# To know or not to know? How looking at payoffs signals selfish

# behavior

Valerio Capraro[1] & Jotte Kuilder[2]

[1]This work was conducted while this author was formally affiliated to the Center for Mathematics and Computer Science (CWI, Amsterdam), but it was conducted outside his research project and it was not financially supported by the CWI.
[2]Institute for Logic, Language, and Computation, Universiteit van Amsterdam, The Netherlands.

**Abstract**

In daily life, people often face a social dilemma in two stages. In Stage 1, they recognize the social dilemma structure of the decision problem at hand (a tension between personal interest and collective interest); in Stage 2, they have to choose between gathering additional information to learn the exact payoffs corresponding to each of the two options or making a choice without looking at the payoffs. While previous theoretical research suggests that the mere act of considering one's strategic options in a social dilemma will be met with distrust, no experimental study has tested this hypothesis yet. What does "looking at payoffs" signal in observers? Do observers' beliefs actually match decision makers' intentions? Experiment 1 shows that the actual action of looking at payoffs signals selfish behavior, but it does not actually mean so. Experiments 2 and 3 show that, when the action of looking at payoffs is replaced by a self-report question asking the extent to which participants look at payoffs in their everyday lives, subjects in *high looking mode* are indeed more selfish than those in *low looking mode*, and this is correctly predicted by observers. These results support Rand and colleagues' Social Heuristics Hypothesis and the novel "cooperate without looking" model by Yoeli, Hoffman, and Nowak. However, Experiment 1 shows that actual looking may lead to different results, possibly caused by the emergence of a moral cleansing effect.

*Keywords*: pro-social behavior, cooperate without looking, envelope game, Prisoner's Dilemma, Dictator Game.

*Word count*: 4347

**Introduction**

Virtually all studies on human pro-sociality assume that decision makers know the exact costs and benefits of a pro-social action beforehand. While this assumption is helpful to develop theoretical models (Fehr & Schmidt, 1999; Bolton & Ockenfels, 2000; Charness & Rabin, 2002; Caprraro, 2013) and conduct behavioral experiments (Rapoport, 1965; Kahneman, Knetsch & Thaler, 1986; Camerer, 2003), in many real life situations people do not actually know the exact payoffs involved beforehand, but can gather this information only in a subsequent stage.

Such situations abound in real life. For example, when a friend asks you to drive her to some store, before making your decision, you can decide to ask for additional information to learn the exact payoff structure of the dilemma (how far is the store? How long should I wait for you?). Similarly, when a friend tells you he is in trouble and needs a temporary loan, before making your decision, you may ask him the exact amount he needs and when he expects to return it. Analogously, before deciding whether to join an ethical cause, you might or might not decide to gather additional information about how much effort (time and money) you need to invest for this cause.

Theorists have started considering these situations only very recently (Hoffman, Yoeli & Nowak, 2015; Hilbe, Hoffman & Nowak, 2015). One simple way to formalize this type of situations is by means of a two-stage strategic game with one decision maker and one observer (see Figure 1). Initially, the decision maker knows that he will have to decide between cooperation and defection. They know that cooperation gives a payoff $d_c$ to themselves and a payoff $o_c = d_c$ to the observer, while defection gives a payoff $d_d > d_c$ to themselves and $o_d < o_c$ to the observer. However, the decision maker does not know the exact payoffs. In Stage 1, the decision maker has to decide between "looking at payoffs" and "not looking at payoffs". In case the decision maker decides to look, she or he learns the complete payoff structure of the game, that is, she or he learns the payoff for both players. Then, in Stage 2, the decision maker makes their actual choice between cooperation and defection.

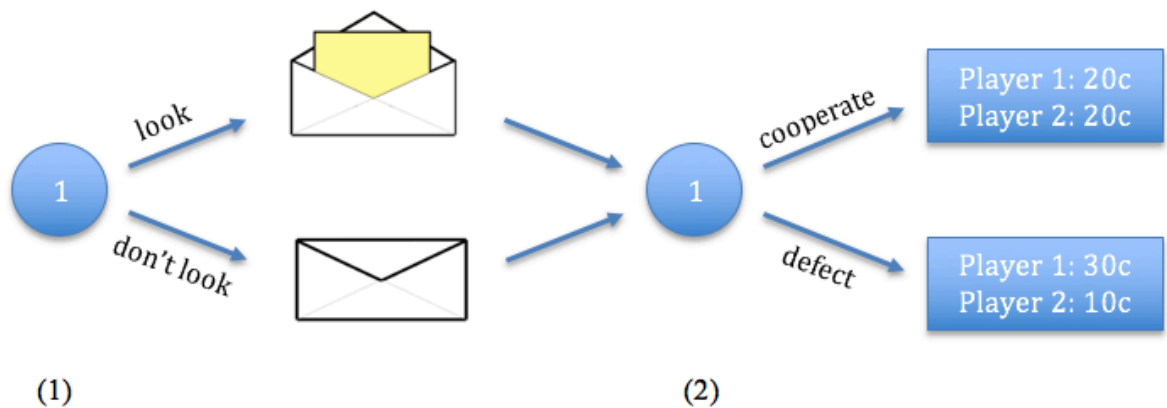

(1)                                                (2)

*Figure 1. Our variant of the envelope game. In stage (1) the decision maker, Player 1, decides whether to look at the payoffs corresponding to cooperation and defection, or not. If Player 1 decides to look at the payoffs, then she or he is informed about the complete payoff structure of the game, that is, she or he learns the payoffs for both players. In stage (2) Player 1 decides whether to cooperate or to defect. Player 2 is passive and has no role. What does looking at the payoffs signal about Player 1's behavior? Do Player 2's beliefs correspond to actual Player 1's intentions?*

This strategic situation is similar to the envelope game introduced for iterated interactions by Hoffman, Yoeli and Nowak (2015) and for one-shot games by Hilbe, Hoffman and Nowak (2015). These theoretical studies posit that "the mere act of considering one's strategic options and gathering information about the possible costs and benefits of an action will be met with distrust" (Hilbe, Hoffman & Nowak, 2015). Indirect theoretical support for this assumption comes from the so-called Social Heuristics Hypothesis (SHH). Introduced by Rand and colleagues (Rand, Greene & Nowak 2012; Rand et al. 2014), the SHH maintains that people internalize strategies that are successful in their everyday interactions and tend to use these heuristics as default strategies when they encounter a new and atypical situation that resembles a situation they have encountered in the past. Then, after additional deliberation, people may overcome these heuristics and adjust their behavior towards the one that is optimal in a given situation.

In our envelope game, the SHH predicts that subjects who look at the payoffs should be more selfish than those to whom payoffs are given by the experimenter. This is due to the fact that gathering information about the exact payoff structure of the game is a signal of deliberation. Moreover, this prediction is in line with the theoretical work by Hoffman, Yoeli and Nowak (2015) and by Hilbe, Hoffman and Nowak (2015).

However, to the best of our knowledge, no experimental studies have been conducted to test this hypothesis. Does looking at payoffs really *signal* selfish behavior in observers? Are people who look at payoffs really more selfish than those making a decision without looking?

## Experiment 1

As mentioned earlier, we aim to (i) measure a possible change in observers' beliefs caused by knowing that the decision maker has decided to look at the payoffs of a social dilemma before making their decision, and (ii) measure whether a possible change in observers' beliefs correspond to a change in decision makers' actual behavior.

### *Method*

Subjects were living in the US at the time of the experiment and were recruited using the online labor market Amazon Mechanical Turk (Paolacci, Chandler & Ipeirotis 2010; Horton, Rand & Zeckhauser 2011, Paolacci & Chandler, 2014). In this and the following studies, we did not conduct an a priori power analysis, but the planned sample sizes were based on previous studies investigating behavioral changes in social dilemma games

(Capraro, Jordan & Rand, 2014).

Each of 1,088 participants (57% males, mean age=32) was randomly assigned to one of seven conditions and passed standard comprehension questions to make sure they have understood the decision problem at hand. Any subjects that did not pass the comprehension questions were automatically excluded from the survey. The seven experimental conditions were as follows.

*Received.* Here all participants were decision makers, to whom we asked to decide

between Option A and Option B. Option A would give 20c to both themselves and the person

they were paired with (participating in the *Guess Received* condition described below). Option

B would give 30c to themselves and 10c to the other person.

*Denied.* This condition was similar to the *Received* condition, but decision makers

(paired with participants in the *Guess Denied* condition) were not told the payoffs

corresponding to the two options. Moreover, participants were not given the choice to learn

them. The only information they had was that Option A would be the fair option and that

Option B would favor themselves at the expenses of the other person.

*Choose.* This condition was similar to the *Denied* condition. Participants were told that

Option A would be the fair option, but Option B would maximize their payoff at the expenses of the other participant. After giving this piece of information, we asked participant whether they wanted to know the exact amounts of money corresponding to each of the two options.

Depending on their choice these participants were paired with observers in the *Guess Chose*

*Yes* and *Guess Chose No* conditions.

*Guess Received.* Each participant was grouped together with other two participants,

named Person 1 and Person 2, who are playing the same game (Person 1 in the role of the decision maker and Person 2 in the role of the receiver). Participants were shown the screenshots of the instructions presented to Person 1 (participating in the *Received* condition) and had to guess Person 1's decision. Correct guesses were incentivized with a $0.20 prize. We opted for measuring beliefs from the point of view of a third party, instead from the point of view of the receiver, because a risk-averse receiver has an incentive not to report their correct beliefs, even if they are incentivized. For example, a risk-averse receiver who believes that the decision maker is going to choose Option A with probability 0.5 and Option B with probability 0.5 would prefer to bet on Option B rather than Option A, because the two bets have the same expected value (20c), but betting on Option B has an higher certain reward. We refer to d'Adda, Capraro and Tavoni (2015) and Dufwenberg and Gneezy (2000) for related discussions.

*Guess Denied.* This condition was similar to the *Guess Received* condition, but participants were shown screenshots of the instructions given to decision makers participating in the *Denied* condition.

*Guess Chose Yes*. This condition was similar to the previous ones, but participants were shown the screenshots of the instructions given to decision makers participating in the *Choose* condition, who decided to look at payoffs. Observers were communicated also the payoff structure of the game, and not only that the decision maker decided to look at the payoffs.

*Guess Chose No*. This condition was similar to the previous ones, but participants were shown the screenshots of the instructions given to decision makers participating in the *Choose* condition, who decided *not* to look at payoffs.

Informed consent was obtained by all participants before the experiment took place, and anonymity was preserved all along the experiment and the analysis of the data.

***Results and discussion***

Figure 2 provides visual evidence of our results. Apart from the condition in which decision makers were given the choice to look at the payoffs and decided to do so, observers' beliefs about decision makers' behavior look very accurate. The only bias seems to regard the condition in which decision makers decided to look at the payoffs. The figure suggests that looking at payoffs signals selfish behavior, but it does not actually mean so.

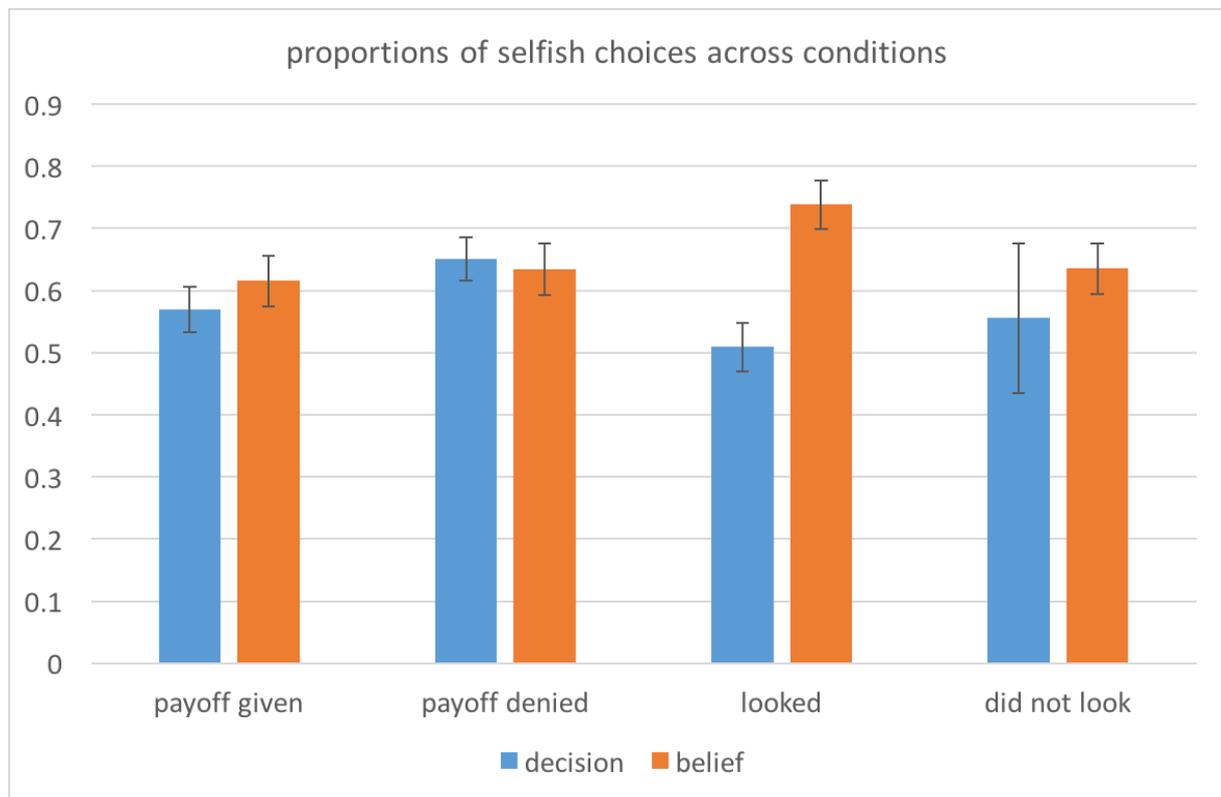

*Figure 2. Proportions of selfish choices across conditions. Error bars denote the standard error of the means. The first, the second, and the fourth pair of columns provide visual evidence that observers' beliefs about decision makers' behavior match actual choices in all conditions, save the one in which decision makers decided to look at the payoffs before making their decision. Logistic regression confirms this (all p's > 0.4). On the other hand, the third pair of columns shows that participants looking at the payoffs were perceived highly*

*more selfish than they actually are. This is confirmed by logistic regression ($\chi^2(1,289) =$ 17.90), coeff = 0.999, z = 3.901, p < .0001, effect size = 23%). This bias turns out to be driven by beliefs, rather than actual behavior. More precisely, participants looking at the payoffs were believed significantly more selfish than in the condition in which payoffs were given ($\chi^2(1,269) = 9.175$, coeff = 0.567, x = 2.130, p = 0.033, effect size = 12.5%), but they were statistically as selfish as those in the baseline ($\chi^2(1,344) = 1.556$, coeff = -0.241, z = 0.217, p = 0.266, effect size = 6%).*

To confirm this intuition, we now report formal statistical analysis. Logistic regression predicting the probability of cooperation as a function of a dummy variable which takes value 1 if a subject participated in the "guess" condition, and 0 otherwise, shows that in the cases of denied payoffs ($\chi^2(1,320) = 3.817$, coeff = -0.071, z = -0.299, p = 0.765, effect size = 2%), received payoffs ($\chi^2(1,324) = 2.123$, coeff = 0.192, z = 0.841, p = 0.400, effect size = 5%) and choosing not to know the payoffs ($\chi^2(1,155) = 1.800$, coeff = 0.330, z = 0.653, p = 0.513, effect size = 7%), there is no statistically significant difference between the choices of decision makers and observers' beliefs. Hence the beliefs about the actions of Player 1 are not significantly different from their actual actions. However, results differ when the decision maker chooses to know the payoffs. Specifically, there is a strong statistically significant difference between the decisions of the participants acting as Player 1 and observers' predictions of their choices ($\chi^2(1,289) = 17.90$), coeff = 0.999, z = 3.901, p < .0001), with an effect size of 23%. Taken together, these results provide evidence for a strong bias according to which subjects looking at the payoffs are perceived much more selfish than they actually are, and this bias is not due to an underlying bias regarding the amount of altruism in others.

Next we examine whether this bias is driven by Player 1's actual altruism or beliefs about Player 1's altruism, or both. To do so, we pool together the data of the *Guess Received* condition and the *Guess Chose Yes* condition and we conduct logistic regression predicting observers' beliefs as a function of a dummy variable, which takes value 1 if the subject participated in the *Guess Chose Yes* condition, and 0 otherwise. Results show that observers' significantly underestimate decision makers' altruism ($\chi^2(1,269) = 9.175$, coeff = 0.567, z = 2.130, p = 0.033), with an effect size of 12.5%. Similarly, we pool together the data of the *Received* condition and those of the participants in the *Choose* condition, who decided to look at the payoffs. Logistic regression shows that participants who looked at the payoffs tend to be more altruist, but not significantly so ($\chi^2(1,344) = 1.556$, coeff = -0.241, z = 0.217, p = 0.266, effect size = 6%). These results provide evidence that the aforementioned bias regarding the level of altruism of subjects who decide to look at the payoffs is mainly driven by a bias in observers' beliefs about decision makers' level of altruism.

Finally, we investigate whether not looking at payoffs signals altruistic behavior. Similar analysis as before shows that these people are neither more altruist ($\chi^2(1,271) = 2.768$, coeff = 0.003, z = 0.012, p = 0.990, effect size = 0.1%) nor perceived to be more altruist ($\chi^2(1,204) = 2.951$, coeff = -0.398, z = -0.798, p = 0.4245, effect size = 9%) than the baseline. However, we mention that the proportion of people who decided not to look at payoffs was so small (around 10%), that it is possible that the lack of a significant effect is due to an undesired ceiling effect.

To summarize, Experiment 1 provides evidence that subjects who look at the payoffs are perceived much more selfish than they actually are.

## Experiment 2

Our first experiment suggests that subjects who look at payoffs are perceived much more selfish than they actually are. One potential explanation for this bias is that the act of looking generates a moral cleansing effect (Sachdeva, Iliev & Medin, 2009). Moral cleansing theory posits that people have a positive moral conception of themselves and that they strive for balance in their moral acts to maintain this positive concept. In other words, when people do something that they think it is morally wrong, they need to subsequently do something that they think it is morally right to compensate for it (Brañas-Garza, Bucheli, Paz Espinosa & García-Muñoz, 2013; Bandura, 1991; Dunning, Fetchenhauer & Schlösser, 2012; Dunning, 2007). As a rather extreme example, it was shown that after contemplating paying the poor to harvest organs, people express an increased desire to donate their own or volunteer for an ideological cause (Tetlock, Kristel, Elson, Green & Lerner, 2000). In this light, it is possible that decision makers are aware of the fact that choosing to know every detail of the decision problem will be perceived with distrust by observers. Since this action then tips the moral balance towards bad behavior, the agent may feel the need to compensate their behavior by cooperating at the next occasion.

To avoid this potentially confounding factor, Experiment 2 replaces the actual act of looking with a self-report question in which subjects are asked the extent to which they try to gather information about the payoff structure of a social dilemma in their everyday interactions, before making a decision. Moreover, in order to better understand whether looking at payoffs is a signal of selfish behavior for *every* subject or, alternatively, there are individual differences according to which looking at payoffs signals selfish behavior for *most* subjects, but for others it signals altruistic behavior, we implement a within-subject design, instead of a between-subject design, as in Experiment 1.

### *Method*

This is a within-subject experiment in which 213 brand new subjects (45% males, average age = 33) participated in the following three conditions, in random order.

In the *looking mode* condition, participants were presented a number of real life situations involving a conflict between one's own benefit and other's benefit (e.g., your friend is in trouble and needs a temporary loan from you. You have to decide between lending them money or not). After presenting the examples, we asked participants the extent to which, in these situations, they try to gather additional information about the exact consequences of their actions, before making a decision. Responses were collected through a 5-point Likert scale from 1 = "very little" to 5 = "very much". After this self-reported question, we measured participants' altruistic attitudes through a standard Dictator Game (DG). In our DG, participants were given 10c and had to decide how much, if any, to give to another anonymous participant (participating in one of the other conditions). The other participant has no active role and only gets what the first player decides to donate. Dictator game donations are usually taken as an individual measure of altruistic attitudes (Engel, 2011) and recent research has shown that they indeed positively correlate with altruism in everyday life (Franzen & Pointner, 2013).

In the *guess no-looking mode* condition, participants were first shown the screenshots of subjects participating in the "looking mode" condition, then told that a participant answered "very little" to our question detecting the looking mode, and finally asked to guess this participant's DG donation towards an anonymous stranger. Correct guesses were incentivized with a 10c reward.

Finally, the *guess yes-looking mode* condition was very similar to the previous one, apart from the fact that subjects were matched with a participant who answered "very much" to our question about looking mode. In reality, to avoid deception, matching between donors and receivers was random, thus all dictators actually donated money.

All participants were asked two comprehension questions to test for their understanding of the decision problem. Participants failing any comprehension questions were automatically excluded from the survey.

### Results and discussion

We start by analyzing whether the measure of looking mode predicts selfish behavior in the Dictator Game. Linear regression predicting DG donation as a function of "looking mode" confirms that this is indeed the case ($F_{(1,211)} = 6.970$, coeff = -0.463 , p = 0.009, $r^2 = 0.032$). Thus, these results confirm the prediction of the SHH that subjects acquiring information about the payoff structure of a social dilemma are more selfish (in a Dictator Game) than those who make a decision without knowing the payoff structure of the social dilemma (see Figure 3).

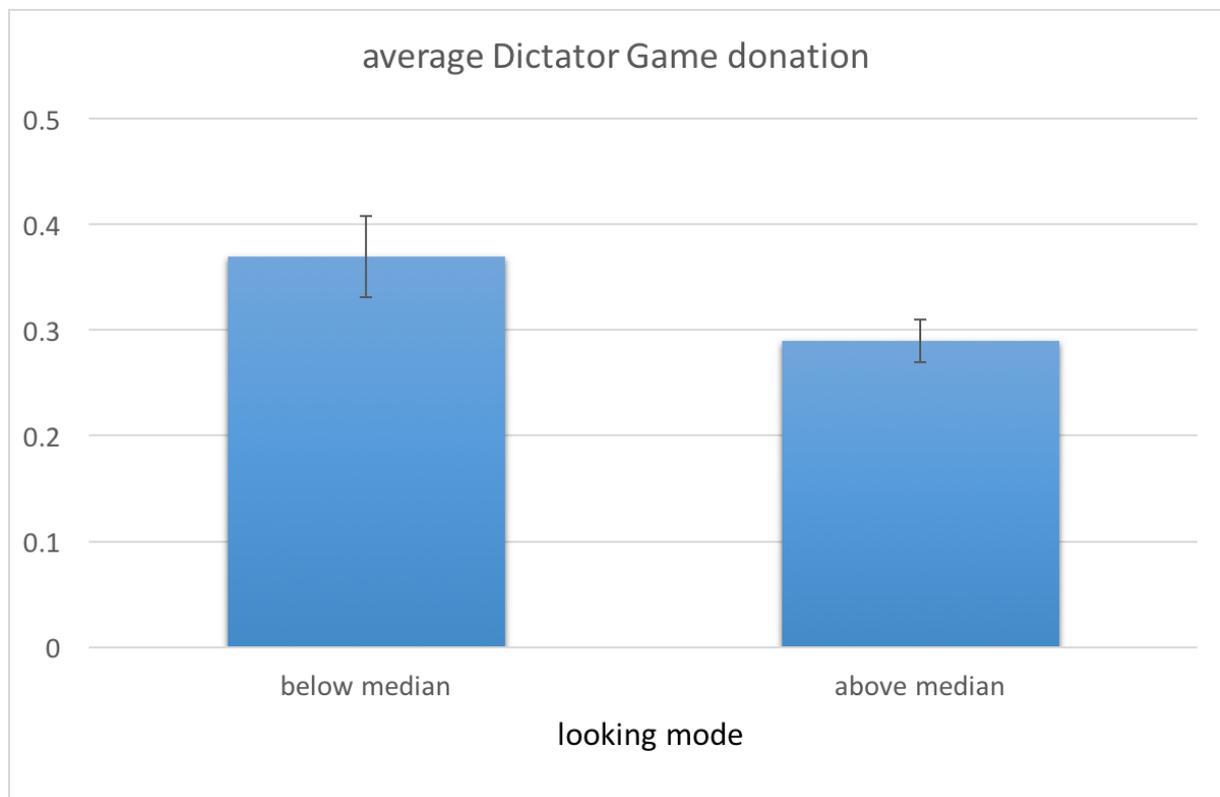

*Figure 3. Average donation in the Dictator Game, broken down by self-reported looking mode (strictly below median vs above median). Looking mode turns out to be a significant predictor of selfish behavior, as confirmed by linear regression (F(1,211) = 6.970, coeff = -0.463, p = 0.009, r$^2$ = 0.032).*

Next we ask whether this behavioral change is correctly predicted by observers. Linear regression predicting observers' choice as a function of a dummy variable that takes value 1 if the observer participated in the *guess no-looking mode* condition and 0 if the observer participated in the *guess yes-looking mode* condition, shows that having an affirmative looking mode is a strong signal of selfish behavior (F(1,414) = 22.767, coeff = 0.141, p < .0001, r$^2$ = 0.052, effect size = 14%). See Figure 4.

Since ours is a within-subject study, this result provides evidence that the same person updates their beliefs when they are paired with a person in looking mode relative to when they are paired with a person in no-looking mode. However, one question remains unsolved: do all subjects update beliefs in the same direction or are there individual differences in the interpretation of looking mode? Interestingly, within-subject analysis shows that 54% of the observers increase their expectation about decision maker's altruism when the decision maker is in a no-looking mode relative to when the decision maker is in a looking mode; 20% of the observers have the same beliefs, regardless of decision maker's looking mode; the remaining 26% of the observers *decrease* their expectation about decision maker's altruism when the decision maker is in a no-looking mode relative to when the decision maker is in a looking mode. Thus, while, on average, looking mode is a signal of selfish behavior, this interpretation is not universal: for a substantial proportion of people, non-looking is a signal of selfish behavior.

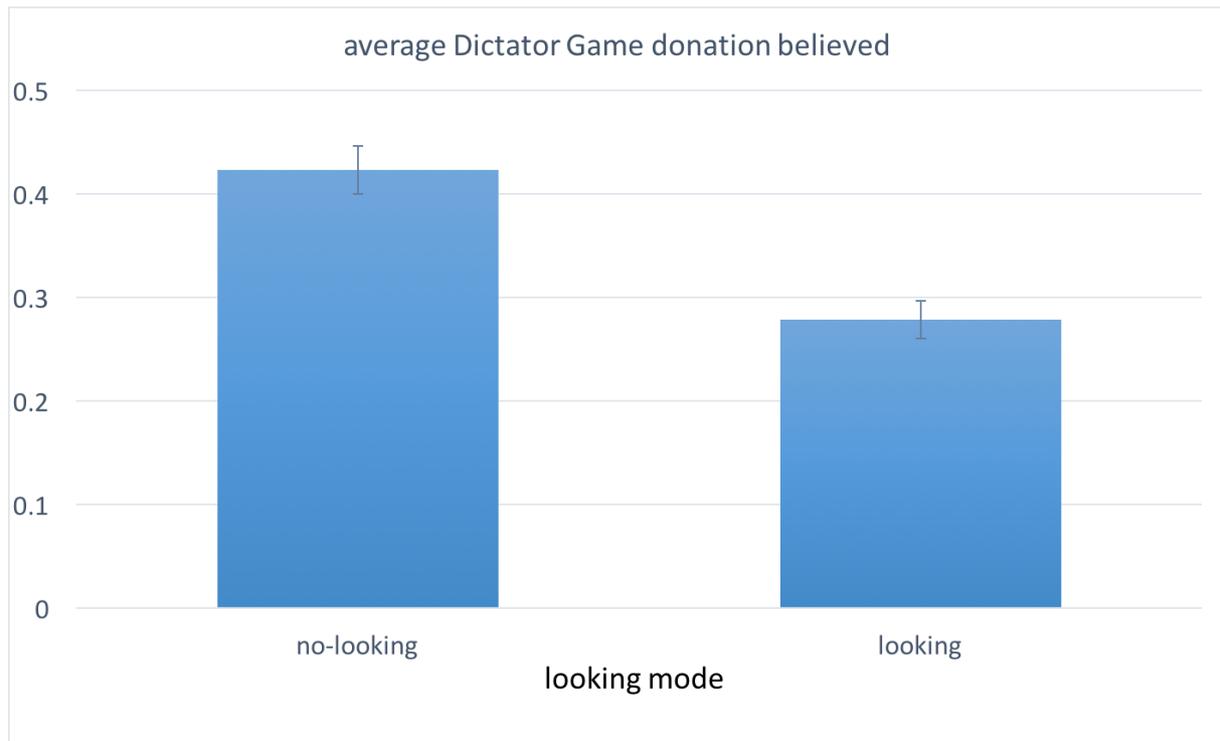

*Figure 4. Average Dictator Game donation believed as a function of Dictator's looking mode. Dictator's looking mode is a strong signal of selfish behavior (F(1,414) = 22.767, coeff = 0.141, p < .0001, r² = 0.052, effect size = 14%). However, within subject analysis shows that, while this negative correlation is true, on average, it is not universally true among all subjects: for 26% of the observers, looking mode is a signal of altruistic behavior, relative to no-looking mode.*

## Experiment 3

Our last experiment aims at extending the findings of Experiment 2 beyond the Dictator game. In the Dictator game, the second player is passive and only receives the amount that the first player decides to give. In the majority of real life situations, however, the second player is not passive and has the opportunity to reciprocate first player's altruistic action. To understand whether our findings extend to this situation, Experiment 3 implements the same design as Experiment 2, but with a Prisoner's Dilemma at the place of the Dictator

Game (we remind that previous research shows that behavior in the Prisoner's Dilemma is *not* equivalent to behavior in the Dictator game: while virtually all subjects who give in the Dictator Game also cooperates in Prisoner's Dilemma, the converse does not hold true. See Capraro, Jordan and Rand, 2014).

### *Method*

Experiment 3 was very similar to Experiment 2. The only difference was that subjects (N = 161, 54% males, average age = 32) played (or were asked to guess how decision makers play) a Prisoner's Dilemma instead of Dictator Game. In our Prisoner's Dilemma, subjects were asked to choose between two options: Option 1 would give 20c to the decision maker and 20c to the other participant; Option 2 would give 30c to the decision maker and 10c to the other person. Participants were told that the other person was given the same set of instructions. We tested participants' understanding of the game through four comprehension questions. Participants failing any of the comprehension questions were automatically excluded from the survey.

### *Results and discussion*

We start by analyzing whether the measure of looking mode predicts selfish behavior in the Prisoner's Dilemma. Logistic regression predicting the probability to cooperate as a function of "looking mode" finds a marginally significant effect ($\chi^2(1,161) = 6.347$, coeff = -0.321, z = 1.960, p = 0.050). Thus, although the correlation is weaker than in Experiment 2, these results confirm the prediction of the SHH that subjects acquiring information about the payoff structure of a social dilemma are more selfish (in a Prisoner's Dilemma) than those who make a decision without knowing the payoff structure of the social dilemma. See Figure 5.

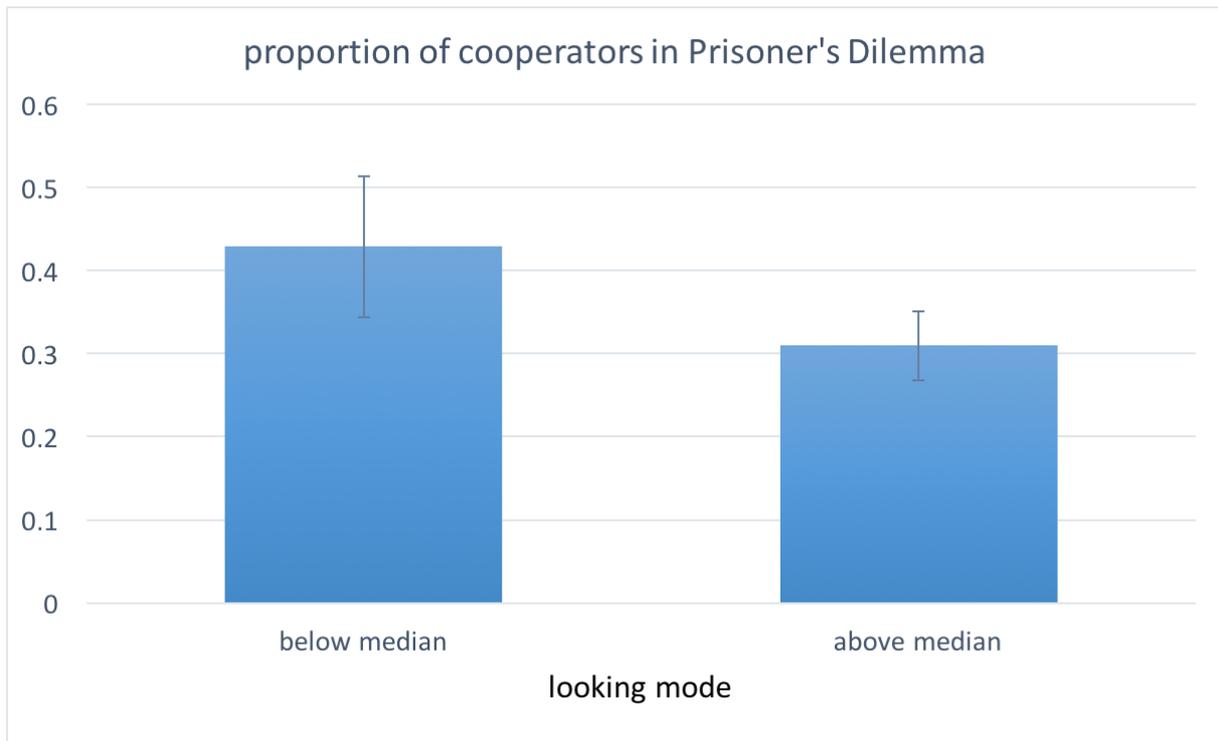

*Figure 5. Average cooperation in the Prisoner's Dilemma, broken down by self-reported looking mode. Looking mode turns out to be a marginally significant predictor of selfish behavior, as confirmed by logistic regression ($\chi^2(1,161) = 6.347$, coeff = -0.321, z = 1.960, p = 0.050).*

Next we ask whether this behavioral change is correctly predicted by observers. Logistic regression predicting observers' choice as a function of a dummy variable that takes value 1 if the observer participated in the *guess yes-looking mode* condition and 0 if the observer participated in the *guess no-looking mode* condition, shows that being in a looking mode is a strong signal of selfish behavior ($\chi^2(1,271) = 7.287$, coeff = -0.694, z = -2.766, p = 0.006, effect size = 17%). See Figure 6.

As in Experiment 2, we finally investigate whether all subjects update their beliefs in the same direction or, alternatively, there are individual differences in the interpretation of looking mode. In line with Experiment 2, within subject analysis shows that 42% of the observers increase their expectation about decision maker's cooperative behavior when the decision maker is in a no-looking mode relative to when the decision maker is in a looking mode; 34% of the observers have the same beliefs, regardless of decision maker's looking mode; and the remaining 24% of the observers *decrease* their expectation about decision maker's cooperative behavior when the decision maker is in a no-looking mode relative to when the decision maker is in a looking mode. Thus, as in Experiment 2, while, on average, looking mode is a signal of selfish behavior, this interpretation is not universal: for a substantial proportion of people, non-looking is a signal of selfish behavior.

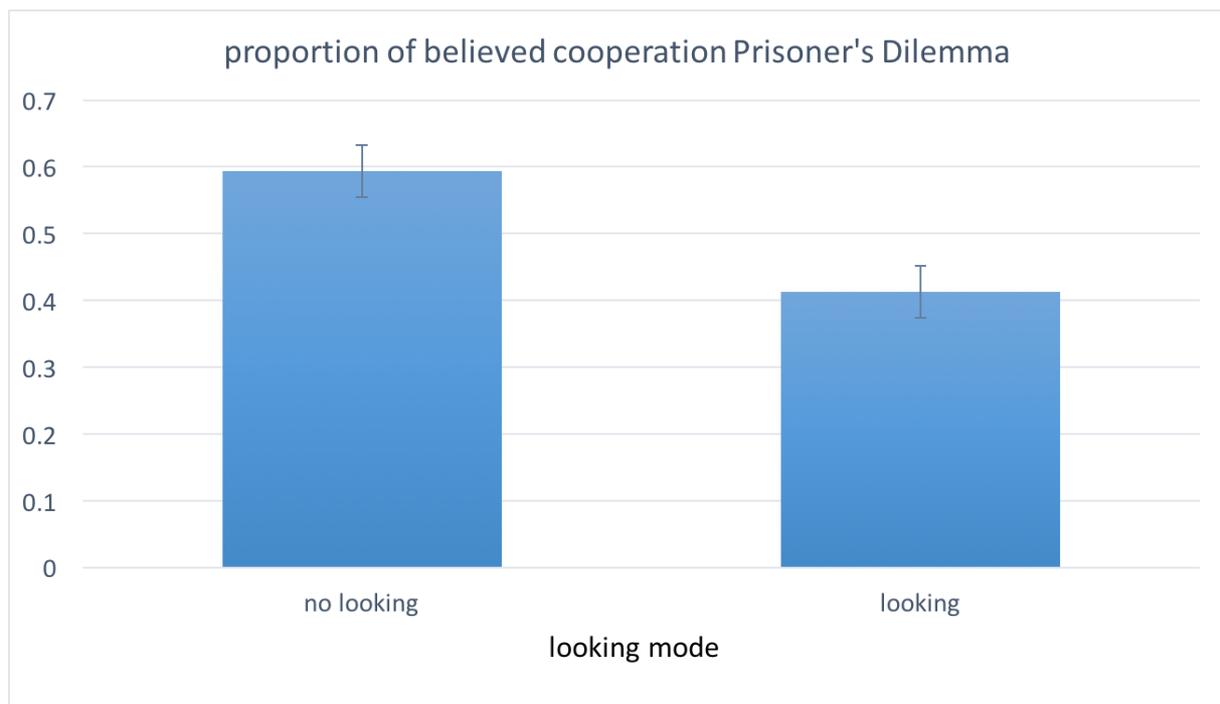

*Figure 6. Average Prisoner's Dilemma cooperation believed by observers' as a function of decision makers' looking mode. Looking mode is a strong signal of selfish behavior ($\chi^2(1,271) = 7.287$, coeff = -0.694, z = -2.766, p = 0.006, effect size = 17%). However, within subject analysis shows that, while this negative correlation is true, on average, it is not*

*universally true among all subjects: for 24% of the observers, looking mode is a signal of non-cooperative behavior, relative to no-looking mode.*

## General discussion

In daily life, people often face social dilemmas in two stages. In Stage 1, they recognize the social dilemma structure of the decision problem (a tension between personal interest and collective interest); in Stage 2, they have to choose between gathering additional information to learn the exact payoffs corresponding to each of the two options or making a choice without looking at the payoffs.

Recent theoretical models propose that looking at the payoffs will be met with distrust (Hoffman, Yoeli & Nowak, 2015; Hilbe, Hoffman & Nowak, 2015). The justification for this assumption is that looking at the payoffs signals deliberative choices rather than intuitive ones, and deliberation has been shown to decrease cooperation in social dilemmas (Rand et al. 2012; Cone & Rand, 2014; Duffy & Smith, 2014; Rand et al. 2014; Lotz, 2015). This decrease occurs particularly among subjects living in a society with high levels of interpersonal trust (Rand & Kraft-Todd, 2014; Capraro & Cococcioni, 2015), for which cooperative heuristics are stronger than among those living in a society with low levels of interpersonal trust.

However, no previous experimental studies have investigated whether this assumption is grounded. Does looking at payoffs really signal selfish behavior in observers? Do observers' beliefs match decision makers' intentions?

Our experiments 2 and 3 provide strong evidence in support of this assumption. Subjects who self-report that, in their everyday life, they generally tend to gather additional information to understand the payoff structure of a social dilemma are both more selfish and perceived to be more selfish than those who self-report that they generally make a decision without collecting additional information about the payoff structure of the social dilemma. In doing so, our results, add to the growing body of literature supporting Rand and colleagues' Social Heuristics Hypothesis (Rand et al., 2012; Rand et al., 2014; Cone & Rand, 2014; Duffy & Smith, 2014; Rand & Kraft-Todd, 2014; Capraro & Cococcioni, 2015; Lotz, 2015; Rand, Newman & Wurzbacher; Peysakhovich & Rand, in press).

Yet, interestingly, there are individual differences in the interpretation of decision maker's looking mode: while about one half of observers believe that looking at the payoffs signals selfish behavior, about one fourth of observers believe the opposite. Since interpreting others' actions in the right way is crucial for healthy and successful social relationships, we believe that understanding the nature of these individual differences in interpreting decision makers' looking mode is an important direction for future research.

On the other hand, Experiment 1 provides the evidence that, while *actual looking* at the payoffs still signals selfish behavior in observers, it is not associated with *actual* selfish behavior. In other words, the act of looking at the payoffs signals selfish behavior, but it does not actually mean so.

What is the origin of this bias? The difference between Experiment 1 and Experiments 2 and 3 was that in Experiment 1 decision makers actually chose to look or not look at the payoffs, while in the other studies we only asked the extent to which subjects generally (i.e., in their everyday life) look at payoffs before making a decision. Hence, we conjecture that this bias may stem from the theory of moral self-concept and moral cleansing proposed by Sachdeva, Iliev and Medin (2009). Their theory posits that people have a positive moral conception of themselves and that they strive for balance in their moral acts to maintain this positive concept. In other words, when people do something that they think it is morally wrong, they need to subsequently do something that they think it is morally right to compensate for it (Brañas-Garza, Bucheli, Paz Espinosa & García-Muñoz, 2013; Bandura, 1991; Dunning, Fetchenhauer & Schlösser, 2012; Dunning, 2007). In this light, it is possible that decision makers are aware of the fact that choosing to know every detail of the decision problem will be perceived with distrust by observers. Since this action then tips the moral balance towards bad behavior, the agent may feel the need to compensate their behavior by cooperating at the next occasion.

Of course, at this stage of research this remains only a conjecture. Other explanations are indeed possible, including, merely, that we found a false negative. Understanding the nature of this bias is certainly another important direction for future research.